\documentclass[3p]{elsarticle}

\usepackage{amsmath}
\usepackage{amssymb}
\usepackage{bbm}
\usepackage[hidelinks]{hyperref}
\usepackage{lineno}
\modulolinenumbers[200]

\journal{Journal of Informetrics}

\biboptions{authoryear}

\begin{document}

\begin{frontmatter}

\title{Comparing scientific and technological impact of biomedical research}

\author{Qing Ke\corref{corrauthor}}
\address{Center for Complex Networks and Systems Research, School of Informatics, Computing, and Engineering,\\Indiana University, Bloomington, Indiana 47408, USA}
\cortext[corrauthor]{Corresponding author}
\ead{qke@indiana.edu}

\begin{abstract}
Traditionally, the number of citations that a scholarly paper receives from
other papers is used as the proxy of its scientific impact. Yet citations can
come from domains outside the scientific community, and one such example is
through patented technologies---paper can be cited by patents, achieving
technological impact. While the scientific impact of papers has been
extensively studied, the technological aspect remains less known in the literature. Here we
aim to fill this gap by presenting a comparative study on how 919 thousand
biomedical papers are cited by U.S. patents and by other papers over time. We
observe a positive correlation between citations from patents and from papers,
but there is little overlap between the two domains in either the most cited
papers, or papers with the most delayed recognition. We also find that the two
types of citations exhibit distinct temporal variations, with patent citations
lagging behind paper citations for a median of 6 years for the majority of
papers. Our work contributes to the understanding of the technological impact of papers.
\end{abstract}

\begin{keyword}
patent-to-paper citation\sep non-patent reference\sep technological impact
\end{keyword}

\end{frontmatter}

\linenumbers

\section{Introduction}

Scientific research builds upon existing knowledge, and such reliance is often
manifested by citing previous scientific papers. Citation flows among papers
therefore have long been used to study the scientific enterprise, such as
mapping knowledge domains~\citep{Rosvall-maps-2008}, tracking the evolution of
science and the emergence of new fields~\citep{Rosvall-mapping-2010,
Sinatra-century-2015}, understanding the formation of scientific
consensus~\citep{Shwed-temporal-2010}, allocating credit in
science~\citep{Shen-collective-2014}, among many others. Citation-based metrics
have been increasingly adopted to assess the scientific impact of various
entities in the scientific community, from papers~\citep{Wang-quantify-2013} and
authors~\citep{Hirsch-index-2005, Sinatra-quantify-2016} to
journals~\citep{Stigler-citation-1994, Varin-statistical-2016},
institutions~\citep{Davis-faculty-1984}, and nations~\citep{King-nation-2004}.
Yet scholarly papers can have their impact that
reaches domains beyond the scientific community. Here, we focus on one such
domain---patented technologies---to study the technological impact of papers,
as patents are the most widely used ones to represent technological
advance~\citep{Fleming-science-2004, Meyer-tracing-2002}. Although such
representation is limited by the possibility that not all patentable inventions
have been patented, scholars have long used patent data to understand
innovative activities and the development of technologies.

Papers that are cited by a patent are listed in the non-patent references
(NPRs) section of a patent application and considered relevant to the
application by either the applicant or the patent examiner. Apart from papers,
many other types of documents may also be listed as NPRs, such as books, Web
pages, etc. Patent law imposes an obligation on patent applicants to submit
relevant ``prior art'' of which they are aware, including both patents and
non-patent materials, and failure to do so may result in the application
unpatentable. Patent examiner who reviews the application may find prior art
themselves, generating additional citations, and then determines the
patentability of the invention.

Studies about the analysis of patent-to-paper citation linkages started already
in the 1980s. \citet{Narin-status-1992} reported a statistical analysis on the types
of NPRs and the time and nation dimension of scientific
NPRs. Scholars have proposed several interpretations about the patent-to-paper
citation linkages. One of the most adopted ones is that the linkage signals
direct knowledge flows~\citep{Jaffe-flow-1993, Azoulay-diffusion-2011}, that is,
the occurrence of citations to a patent or a paper is argued to indicate that
the inventors have benefited from the patent or the paper. This interpretation,
although subject to debate~\citep{Meyer-does-2000} and limited by the fact that
patent examiners can also add citations~\citep{Alcacer-examiner-2006,
Alcacer-overview-2009, Lemley-examiner-2012}, has been the basis of many
studies that attempt to demonstrate how publicly-financed research contributes
to technological advances and private-sector
innovations~\citep{Narin-linkage-1997, McMillan-biotech-2000,
Ahmadpoor-dual-2017, Azoulay-public-2017, Li-applied-2017}, motivating further
public support for scientific research. \citet{Fleming-science-2004} argued that the
mechanism through which science increases the rate of invention is that science
leads inventors' search process more directly to useful
combinations. Other scholars have argued that
patent-to-paper citations signal relatedness between science and
technology~\citep{Callaert-source-2014}. 

Our work shifts the attention from interpreting the linkage to assessing the
technological impact of papers. In this regard, a related line of literature is
the examination of the broader impact of research beyond the traditional
scientific community. Existing work has examined how papers are covered by news
media~\citep{Phillips-lay-1991}, used in the development of drug
products~\citep{Williams-from-2015}, referenced in clinical
guidelines~\citep{Grant-clinical-2000}, and mentioned on the social Web
(e.g., Twitter and Wikipedia)~\citep{Thelwall-altmetrics-2013}, among many other
outlets. Our work extends this line of literature by focusing on the
technological community, which hitherto has been less explored, and
investigates how papers are cited by patents. Moreover, we emphasize citation
growth over time rather than simple citation counts at a particular time point,
allowing us to explore the dynamics of the utilization of scientific research
for technology development. The only study that is similar to our work is the
one by~\citet{Ding-explore-2017}. However, we not only look at the
entire citation history of papers as opposed to two five-year time windows
considered by them, but also make a comparison of citations received from
patents and from papers.

Based on the cohort of biomedical papers that have received citations from U.S.
patents, we perform a comparative study on how they are cited by patents and by
other papers over time. We report a set of stylized facts about the two types of
citations. First, similar to paper citations, patent citations are also
heterogeneously distributed. Yet, highly-cited papers in the two domains have
small overlap. Second, there are delayed-recognition papers that achieve high
popularity among patents after years of dormant. Third, patent citations
generally lag behind paper citations for the majority of papers.

\section{Data and methods}

For each paper, we assembled two types of citations, namely those from other
papers and from patents. We first describe the patent citation case. We focused
only on U.S. patents, due to the public availability of patent bibliographic
data over a long period of time. To get citation information from patents, we
downloaded the front page bibliographic data of all utility patents granted by
the United States Patent and Trademark Office (USPTO) between $1976$ and $2013$
from \url{https://bulkdata.uspto.gov}, and excluded withdrawn
patents~(\url{https://www.uspto.gov/patents-application-process/patent-search/withdrawn-patent-numbers}).
Each patent included a list of references containing previously issued patents
and optional NPRs that can refer to essentially anything, including books,
papers, patent applications, online resources, etc. As we are interested in
papers, we further excluded patents without any NPR, ending up with $1,637,072$
patents.

We then matched their NPRs with papers indexed in MEDLINE, a
large-scale bibliographic database for biomedical research literature
maintained by the National Library of Medicine (NLM). The matching steps are
as follows:
\begin{enumerate}
\item We submitted a search query to PubMed where the search term is the entire
NPR text, with the URL following:
\url{https://www.ncbi.nlm.nih.gov/pubmed/?term=[NPR]&report=uilist&format=text},
which returned the matched PMID.
\item If the first step failed, we then extracted relevant bibliographic
information, such as author, title, journal, volume, number, pages, and year,
available in the NPR text, using the \texttt{AnyStyle} parser
(\url{https://anystyle.io/}). It employs a machine learning technique called
Conditional Random Fields to parse citation text with any style, which is
achieved by pre-training the model with different styles of labeled citation
text.
\item After the extraction, we searched PubMed using the \texttt{ECitMatch}
E-utility (\url{https://dataguide.nlm.nih.gov/eutilities/utilities.html\#ecitmatch}),
which accepts 5 types of information as input, namely journal, year, volume,
first page, and author name, and returns the matched PMID. When searching, we
first used all the 5 fields and, if failed, all the 5 possible combinations of
4 fields.
\end{enumerate}
To validate our matching results, we manually matched $208$ randomly selected
NPRs, and Table~\ref{tab:match-conf-mtrx} reports the confusion matrix. For
$117$ cases, hand labeling and the matching method found the same paper (true
positives); for $85$ cases, both agreed that there was no paper matched (true
negatives); and for $6$ cases, our matching method deemed that there was no
paper matched but manual labeling found one (false negatives). A total of
$919,222$ unique papers were matched.

\begin{table}[t]
\centering
\caption{Confusion matrix for results of matching non-patent references to
MEDLINE papers.}
\label{tab:match-conf-mtrx}
\begin{tabular}{|l | l|}
\hline
True positive = 117 & False positive = 0 \\
\hline
False negative = 6 & True negative = 85 \\
\hline
\end{tabular}
\end{table}

After this integration step, we counted, for a paper published in calendar year
$t_p$, the number of patent citations it received in the $t$-th
($t \in [0, L]$; $L = T - t_p$; $T = 2013$) year after its publication, denoted as
$c_t^P$. Note that the citing patents in each year $t$ (calender year $t_p + t$) are
those \emph{issued} at time $t$ rather than those whose applications were
submitted at $t$. The total number of patent citations it
received until the end of the observation period is
$C^P = \sum_{t=0}^{T- t_p} c_t^P$. Here we do not distinguish between citations
generated by applicants or examiners, as we do not concern about knowledge
spillover and examiner-added citations may also indicate the impact of papers.

To get the number of citations from papers, we turn to the Web of Science (WoS) database, as
citation data is available only for PubMed Central papers. We used a version of
WoS currently housed at the Indiana University Network Science Institute to
retrieve the paper citation data. To locate MEDLINE papers in WoS, we used the
mapping data between PMID (PubMed ID) and UT (Accession Number), which are the
two identifiers used in their respective database, and successfully found
$859,085$ ($93.46\%$) MEDLINE papers in WoS. When counting citations, we only
considered the following types of documents: article, review, editorial, note,
and letter. For each of the papers under consideration, we denote its yearly
number of citations from other papers as $c_t^A$, and $C^A$ is the total number
of paper citations it received by $2013$.

For analysis that involved the entire MEDLINE database, we used a snapshot in
$2015$ that contained $23,343,329$ papers. To get the fields of papers, we
chose to use the NLM Catalog data
(\url{https://www.ncbi.nlm.nih.gov/nlmcatalog}), as it is specifically created
to be used in conjunction with other databases such as MEDLINE maintained by
NLM. It assigns each indexed journal to one or more categories called Broad
Subject Terms (e.g., Biochemistry, Cell Biology, Nursing, Health Services
Research).

\section{Results}

The overarching goal of the present study is to examine how patent citations of
papers are different from paper citations. To this end, we present four sets of
results. First, we report in Section~\ref{subsec:field} descriptive statistics
on the types of papers that get cited by patents, journals where these papers
were published, and fields to which they belong, given that these statistics
are not well-known in the existing literature. Second, in
Section~\ref{subsec:total}, we examine how total patent and paper citations
differ. Next, Section~\ref{subsec:sb} examines how patent and paper citations
change over time by focusing on the delayed recognition phenomenon. Finally,
Section~\ref{subsec:ts} performs a lead-lag analysis of citation dynamics.

\subsection{Fields and journals} \label{subsec:field}

The $919,222$ papers that get patent citations only account for a very small
portion ($4\%$) of all $22,975,980$ MEDLINE papers published until $2013$. Our
estimation is similar to the one obtained from a recent study where papers in
WoS rather than MEDLINE were considered: $1.41$ out of $32$ million, or $4.4\%$
of, WoS papers were cited by USPTO-issued patents~\citep{Ahmadpoor-dual-2017}.

Table~\ref{tab:doc-type} shows the distribution of document types assigned by
WoS for the $859,085$ MEDLINE papers indexed there. Articles contribute to the
largest portion ($86.6\%$), followed by review, note, editorial, and letter.

\begin{table}[t]
\centering
\caption{Document type distribution for $859,085$ papers cited by USPTO-issued
patents.}
\label{tab:doc-type}
\begin{tabular}{| l | r | r |}
\hline
Type & Count & $\%$\\
\hline
Article & $744,309$ & $86.64$ \\
Review & $63,709$ & $7.42$\\
Note & $29,344$ & $3.42$ \\
Editorial & $10,520$ & $1.22$ \\
Letter & $8,544$ & $0.99$ \\
Others & $2,659$ & $0.31$ \\
\hline
\end{tabular}
\end{table}

Table~\ref{tab:top-cat} displays the top $20$ most cited biomedical research
fields, as defined by NLM as Broad Subject Terms, which in total account for
$73.1\%$ of patent citations. For each field, we derived
three statistics: (1) the total number of patent citations of all papers
published in journals that belong to the field, (2) the unique number of papers
that are cited by patents, and (3) the fraction of papers that are cited by
patents among all papers published there. While the first two are of
retrospective, the third one is of prospective, since each journal publishes
different amounts of papers. Biochemistry is the most cited field, attracting
$12.9\%$ of citations from patents, followed by Science ($10.6\%$), Molecular
Biology ($6.0\%$), Allergy and Immunology ($5.2\%$), Cell Biology ($4.2\%$),
and Chemistry ($3.7\%$). Here ``Science'' covers multi-disciplinary journals
like \emph{Nature} and \emph{Science}, similar to the Multidisciplinary Science
designated in Journal Citation Reports (JCR).

\begin{table*}[t]
\centering
\caption{List of top $20$ most cited fields. For each field, as defined by NLM
as Broad Subject Term, we counted the total number of patent citations of
papers published in journals that belong to the field, as well as the unique
number of papers cited by patents and the fraction of these papers among all
papers in this field. Journals that are designated to multiple fields are
counted multiple times. ``Science'' covers multi-disciplinary journals.}
\label{tab:top-cat}
\begin{tabular}{l r r r r r c}
\hline
Field & \# Cites & \% Cites & \# Papers & \% Papers & \% Published & Category \\ \hline
Biochemistry & $603,322$ & 12.86 & $143,381$ & 12.26 & 11.58 & Basic \\
Science & $495,804$ & 10.57 & $57,054$ & 4.88 & 9.64 & -- \\
Molecular Biology & $282,126$ & 6.02 & $62,826$ & 5.37 & 11.19 & Basic \\
Allergy and Immunology & $243,044$ & 5.18 & $58,866$ & 5.03 & 11.09 & Clinical \\
Cell Biology & $198,189$ & 4.23 & $41,998$ & 3.59 & 10.30 & Basic \\
Chemistry & $175,537$ & 3.74 & $44,047$ & 3.77 & 7.96 & -- \\
Pharmacology & $167,290$ & 3.57 & $46,249$ & 3.96 & 7.11 & Clinical \\
Neoplasms & $162,979$ & 3.48 & $45,752$ & 3.91 & 6.64 & -- \\
Medicine & $158,470$ & 3.38 & $38,109$ & 3.26 & 1.69 & Clinical \\
Biotechnology & $127,242$ & 2.71 & $22,701$ & 1.94 & 14.04 & -- \\
Neurology & $101,226$ & 2.16 & $32,322$ & 2.76 & 3.62 & Clinical \\
Biophysics & $97,713$ & 2.08 & $27,627$ & 2.36 & 6.62 & Basic \\
Virology & $83,114$ & 1.77 & $20,188$ & 1.73 & 13.83 & Basic \\
Physiology & $82,901$ & 1.77 & $25,268$ & 2.16 & 4.25 & Basic \\
Cardiology & $82,898$ & 1.77 & $19,875$ & 1.70 & 3.61 & -- \\
Microbiology & $78,376$ & 1.67 & $25,299$ & 2.16 & 6.78 & Basic \\
Ophthalmology & $75,156$ & 1.60 & $22,185$ & 1.90 & 5.74 & Clinical \\
Vascular Diseases & $71,664$ & 1.53 & $19,435$ & 1.66 & 5.14 & -- \\
Biology & $70,833$ & 1.51 & $19,696$ & 1.68 & 4.19 & -- \\
General Surgery & $68,991$ & 1.47 & $15,409$ & 1.32 & 1.80 & Clinical \\
\hline
\end{tabular}
\end{table*}

The second observation from Table~\ref{tab:top-cat} is that the share of
citations for each field is roughly proportional to the share of the number of
unique papers cited, except for the Science category. This means that
multidisciplinary journals accrue patent citations disproportionately,
suggesting a larger-than-average number of patent citations for papers
published there.

When looking at the fraction of cited papers among all papers published (``\%
Published'' column), we observe that the overall tendency to be cited by
patents varies across fields. Among these most cited fields, Biotechnology has
the largest portion ($14\%$) of papers cited by patents. Virology,
Biochemistry, Molecular Biology, Allergy and Immunology, and Cell Biology all
have more than $10\%$ of such papers. On the other hand, only $1.7\%$ of papers
belonging to Medicine get patent citations, which is similar for General
Surgery. Cardiology, Neurology, Biology, and Physiology also generate a small
fraction of patent-cited papers.

The last column in Table~\ref{tab:top-cat} indicates whether these fields
belong to basic research or clinical medicine, as categorized by Narin et~al.
in the $1976$ pioneering work on the structure of biomedical
literature~\citep{Narin-bio-1976}. Although the number of basic research fields
is similar to the ones belonging to clinical medicine, basic research surpass
clinical medicine once we weight by total citations or unique papers. This
resonates with previous results~\citep{Narin-linkage-1997, McMillan-biotech-2000}.

Next, we delve into journals. Table~\ref{tab:top-jnl} reports the top $10$
field-specific journals that received the most patent citations. For each
journal, we present the same set of statistics as in Table~\ref{tab:top-cat}.
We see from Table~\ref{tab:top-jnl} that papers that obtained patent citations
were published in leading journals. Across fields, \emph{PNAS},
\emph{Journal of Biological Chemistry} (JBC), a journal with a long publishing
history, and \emph{Science} are the top three most cited journals. They are
also the three journals that published the largest number of papers that are
cited by patents. Other highly-cited journals include \emph{Nature},
\emph{Journal of Medicinal Chemistry}, \emph{Nucleic Acids Research}, and
\emph{Cell}.

Similar to what has been observed in Table~\ref{tab:top-cat}, most journals
attract patent citations proportionate to their share of cited papers. But this
is not the case for \emph{Cell}: only $15\%$ of Cell Biology papers were
published in \emph{Cell}, yet they account for $31\%$ of citations to the
field. Other prominent examples, although to a lesser extent, include
\emph{Nucleic Acids Research}, \emph{Journal of Medicinal Chemistry},
\emph{Nature}, and \emph{Science}.

Table~\ref{tab:top-jnl} also provides results from a prospective analysis.
Across these fields, \emph{Annual Review of Immunology} has the highest
fraction ($54\%$) of papers that are cited by patents, followed by
\emph{Journal of Medicinal Chemistry} ($43\%$), \emph{Cell} ($38\%$), and
\emph{EMBO Journal} ($30\%$). On the other hand, \emph{Nature} and
\emph{Science} have a relatively low fraction, which could simply due to the
fact that they are multidisciplinary journals that publish non-biomedical
papers.

A final point regarding both Tables~\ref{tab:top-cat} and \ref{tab:top-jnl} is
that these results are limited by the fact that MEDLINE is a database for the
biomedical research literature. As such, it may have low coverage of papers in
other disciplines such as physics and engineering, especially if papers in
these disciplines are not directly related to biomedicine. This may to some
extent dictate our results. For example, in a seminal work by \citet{Narin-linkage-1997} that
studied citations from U.S. patents to papers, they
found that \emph{Tetrahedron} is among the top most cited Chemistry journals.
However, it fails to make it top in Table~\ref{tab:top-jnl}, because MEDLINE
only has a limited coverage of papers in \emph{Tetrahedron}. Future work
therefore is needed to compare how our results are different from the ones
based on other databases like WoS.

\subsection{Total citations} \label{subsec:total}

How many citations does a paper receive from patents? How does it compare to
citations from other papers? In this section, we investigate total citations.
Tables~\ref{tab:top-ptc} and \ref{tab:top-ppc} list the top $10$ papers by
total patent citations $C^P$ and paper citations $C^A$, respectively. The paper
with the highest number of paper citations in our sample happens to be the most
cited one of all time~\citep{Noorden-top-2014}.

\begin{figure*}[t]
\centering
\includegraphics[trim=0mm 5mm 0mm 0mm, width=\textwidth]{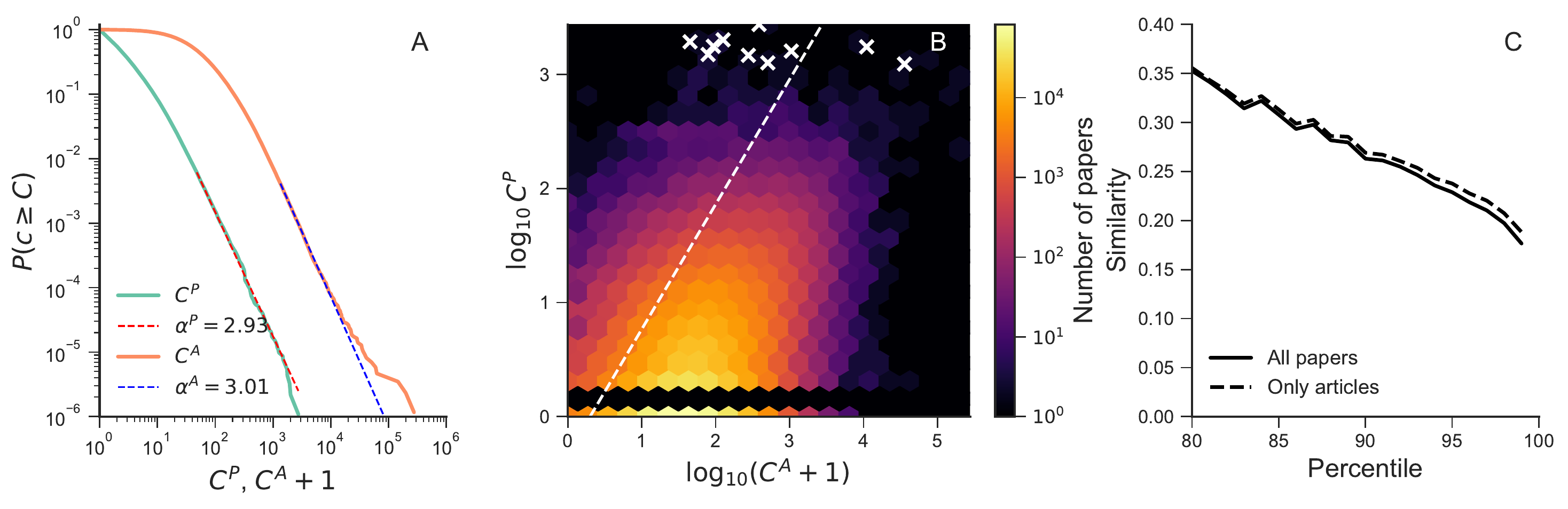}
\caption{Patent and paper citation statistics. (A) Survival distribution
functions of total number of patent citations $C^P$ and paper citations $C^A$
across all papers that are cited by at least one U.S. patent ($C^P \geq 1$).
Both $C^P$ and $C^A$ are measured until $2013$. The red dashed line corresponds
to the estimation of the power-law distribution
$p(C^P) = \frac{\alpha^P-1}{C_{\min}^P} \left( \frac{C^P}{C_{\min}^P} \right)^{{-\alpha}^P}$,
where $C_{\min}^P = 45$ and $\alpha^P = 2.93$. The blue dashed line is the
power-law fit of $C^A$, where $C_{\min}^A = 1344$ and $\alpha^A = 3.01$. Both
are estimated using the method developed in \citet{Alstott-powerlaw-2014} and
\citet{Clauset-powerlaw-2009}. (B) Heat map between $\log_{10} (C^A+1)$ and
$\log_{10} C^P$. The color encodes the number of papers. The white dashed line
corresponds to $C^P = C^A$. The white crosses highlight the top $10$ most cited
papers by patents (Table~\ref{tab:top-ptc}). (C) Overlap between the two
lists of top cited papers by $C^P$ and $C^A$.}
\label{fig:c-total}
\end{figure*}

Heterogeneity in the number of patent citations $C^P$ is present for the cohort
of $919,222$ papers that got cited by patents, as evidenced from
Fig.~\ref{fig:c-total}A where we plot the survival distribution of $C^P$.
Although $414,981$ ($45.1\%$) papers have only one patent citation, there
exists papers that are cited by thousands of patents. We fitted the
distribution with the power-law function, giving us exponent $\alpha^P = 2.93$.
For comparison, we also show in Fig.~\ref{fig:c-total}A the distribution of
$C^A$, the total number of paper citations, and the power-law fit yields
exponent $\alpha^A = 3.01$. This suggests that despite paper citations are
larger than patent citations---which could simply due to the fact that there is
a larger pool of papers than that of patents---they exhibit similar speed of
decay.

We further compare total patent and paper citations.
Fig.~\ref{fig:c-total}B plots $C^P$ against $C^A$ in the form of heat
map, where the color encodes the frequency of papers. The map features a broad
band with an upward slope, indicating that papers under our consideration that
have more paper citations in general tend to attract more patent citations as
well. The Spearman's rank correlation coefficient between $C^P$ and $C^A$ is
$0.228$. Such a positive correlation persists (coefficient $0.251$) if we
consider all the $15,678,754$ MEDLINE papers that can be found in WoS. We also
observe that the region of the highest density is located in
the lower left of the heat map, corresponding to the case where most papers
have paper citations less than $100$ and patent citations less than $10$. The
map indicates that the vast majority of papers have less patent citations than
paper citations, but $22,736$ ($2.65\%$) papers exhibit the opposite, among
which $8$ of the top $10$ papers with most patent citations are in this case.
A total of $6,166$ ($0.7\%$) papers are cited by patents but got zero paper citation.

Are papers that are highly cited in the scientific community also highly cited
in the patent sphere? We quantified the extent of overlap between the two sets
of top cited papers using a similarity measure. Formally, let $M^P$ ($M^A$) be
the set of papers with the number of patent (paper) citations no less than the
threshold corresponding to a given percentile, and the similarity between $M^P$
and $M^A$ is defined as
$s = \left\vert M^P \cap M^A \right\vert / \left\vert M^P \right\vert$, which
measures the fraction of top cited papers by patents that are also top cited by
papers. Fig.~\ref{fig:c-total}C shows that similarity $s$ steadily
decreases as we increase the percentile. Only $18\%$ of the top $1\%$ most
cited papers by patents are also in the top $1\%$ by the number of paper
citations, indicating a small overlap of papers that are highly cited in
both the scientific and the technology community. This pattern is consistent if we
consider only research articles (Fig.~\ref{fig:c-total}C) or papers in one filed
(Fig.~\ref{fig:top-sim}A).

\subsection{Delayed recognition papers} \label{subsec:sb}

The previous section has examined the total number of citations measured at the
end of our observation period $T = 2013$, but how it reached to that number can
be diverse. We now look at time-dependent citation growth. From now on, we only
restrict our analysis to the cohort of $852,919$ papers that (1) were published
from $1976$ and onward, since patent citation data is available only starting
from $1976$, and (2) had at least one paper citation ($C^A > 0$), since we are
interested in the comparison between $c_t^P$ and $c_t^A$.

We first focus on a class of papers---the so-called ``Sleeping Beauty'' papers
that lie dormant in a long period of time after their publication and then
suddenly become highly cited. This notion has been mostly constrained within
the scientific community, that is, citation curves based on which SBs are
discovered are derived from how many other scientific papers have cited the
focal one. Recent work has started to extend this notion to the technology
domain~\citep{vanRaan-sb-2017}, and here we investigate whether there are also
SBs that are perceived as late boomer by the technology community.

\begin{figure}[t]
\centering
\includegraphics[trim=0mm 5mm 0mm 0mm, width=\columnwidth]{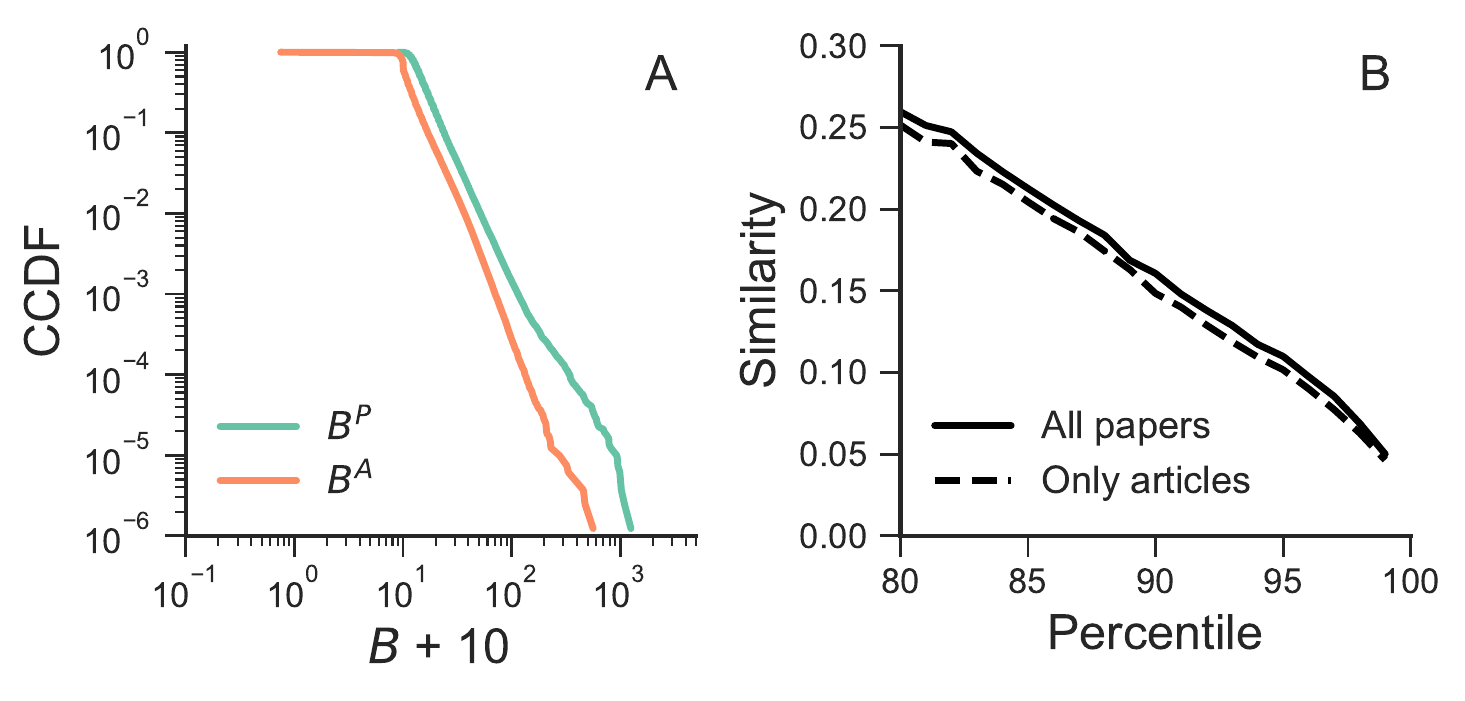}
\caption{Beauty Coefficient $B^P$ and $B^A$ calculated based on patent and
paper citation dynamics. (A) Distributions of $B^P$ and $B^A$. (B) Similarity
of the sets of top SBs based on different percentiles.}
\label{fig:sb}
\end{figure}

To do so, we calculated the Beauty Coefficient based on the $c_t^P$ and $c_t^A$
curve of each paper~\citep{Ke-SB-2015}, denoted as $B^P$ and $B^A$,
respectively. Tables~\ref{tab:top-sb-pt} and \ref{tab:top-sb-pp} report the top
$10$ SBs by $B^P$ and $B^A$, respectively. Fig.~\ref{fig:sb}A, which
plots the distributions of $B^P$ and $B^A$, indicates that the extent of
delayed recognition of papers perceived by both the scientific and technology
community spans three orders of magnitude, similar to what has been observed
before~\citep{Ke-SB-2015}. Regarding individual papers, there is a negligible
correlation between their $B^P$ and $B^A$ values (Spearman correlation
coefficient $0.09$). There is a low overlap of top SBs recognized by the two
communities, as shown in Fig.~\ref{fig:sb}B: only $5\%$ of the top $1\%$
SBs measured from patent citations also rank in the top $1\%$ SBs measured from
paper citations. This observation still holds even if we consider only articles
(Fig.~\ref{fig:sb}B) or papers in one field (Fig.~\ref{fig:top-sim}B). These
results suggest different life-cycles of patent and paper citations.

\subsection{Time-dependent citation accumulation} \label{subsec:ts}

Having looked at a particular type of papers, we now characterize the temporal
variation of citations. To do so, we introduce the following parameters to
describe a given citation dynamics curve $c_t$:
\begin{enumerate}
\item $t_f = \arg \; \left\{ \min_{t} \, \sum_{t' = 0}^t c_t >0 \right\}$. It
measures the number of years taken to obtain the first citation;
\item $t_m = \arg \; \left\{ \max_{t} \, c_t \right\}$. It is the number of
years taken to obtain the maximum yearly citations;
\item $I = \mathbbm{1} \left( \exists t \text{ s.t. } c_t < c_{t_m}/2, t \in [t_m+1, L] \right)$.
It indicates whether the yearly citations have decreased to half of the maximum;
\item $\tau = \begin{cases} L - t_m, & I = 0 \cr t_h - t_m, & I = 1 \end{cases}$.
The first case captures the number of years that the curve has stayed above
$c_{t_m} / 2$ after reaching its maximum, given that the curve has not fallen
below half of the maximum. The second case measures the number of years taken
to fall below the half of the maximum, where $t_h$ is the time when $c_t$ drops
below $c_{t_m}/2$ for the first time.
\end{enumerate}
While the first two summarize how $c_t$ reaches the maximum, the latter two
characterize how $c_t$ decrease after that. As each paper is associated with
two time-series, namely $c_t^P$ and $c_t^A$, we further compute two additional
parameters: $\Delta t_f = t_f^P - t_f^A$ and $\Delta t_m = t_m^P - t_m^A$,
capturing how many years the first and the maximum patent citation lag behind
the paper citation case.

\begin{figure}[t]
\centering
\includegraphics[trim=0mm 5mm 0mm 0mm, width=\columnwidth]{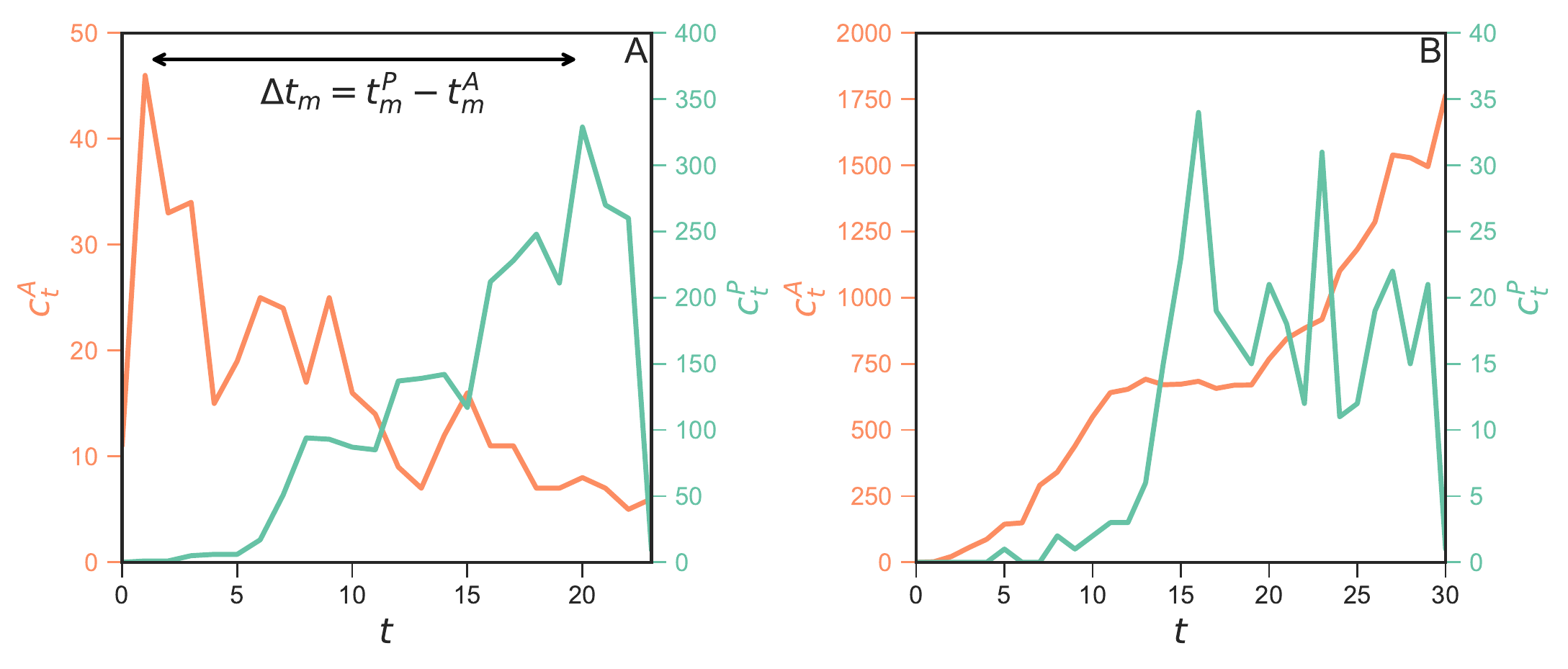}
\caption{Yearly number of paper citations $c_t^A$ (left axis) and patent
citations $c_t^P$ (right axis) of (A) \citet{Bowie-decipher-1990} and (B)
\citet{Mosmann-rapid-1983}.}
\label{fig:ct-ex}
\end{figure}

As illustration, Fig.~\ref{fig:ct-ex} shows $c_t^P$ and $c_t^A$ for two papers.
For the first paper~\citep{Bowie-decipher-1990}, which was published in $1990$,
its yearly paper citations $c_t^A$ reached its maximum quickly ($t_f^A = 0$,
$t_m^A = 1$) and then faded away steadily ($I_A = 1$, $\tau_A = 3$), whereas
the patent citations $c_t^P$ kept increasing for $20$ years ($t_f^P = 1$,
$t_m^P = 20$) and then quickly died out ($I_P = 1$, $\tau_P = 3$), therefore
$\Delta t_f = 1$ and $\Delta t_m = 19$. For the second
paper~\citep{Mosmann-rapid-1983}, published in $1983$, its yearly paper
citations reached the peak at the end of the observation period ($t_f^A = 1$,
$t_m^A = 30, I^A = 0, \tau^A = 0$), whereas the patent citations climbed to the
peak $16$ years after publication, yielding $\Delta t_m = -14$.

\begin{figure}[t!]
\centering
\includegraphics[trim=0mm 5mm 0mm 0mm, width=\columnwidth]{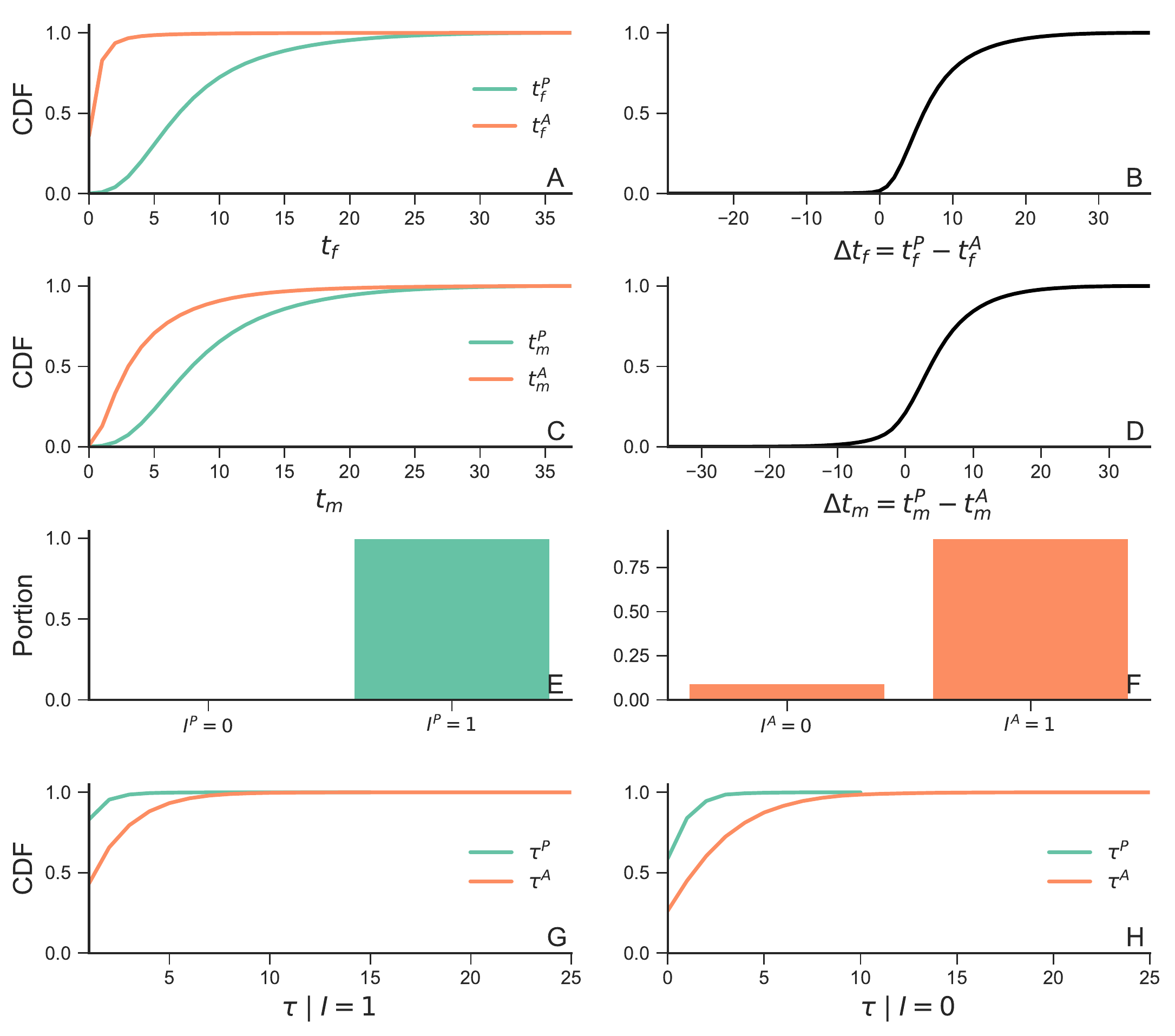}
\caption{Distribution of parameters characterizing the temporal variation of
patent and paper citations. (A) Cumulative distribution function of the
number of years taken to get the first patent ($t_f^P$) and paper ($t_f^A$)
citation. (B) CDF of $\Delta t_f = t_f^P - t_f^A$. (C) CDF of the
number of years taken to reach to the maximum yearly patent ($t_m^P$) and paper
($t_m^P$) citations. (D) CDF of $\Delta t_m = t_m^P - t_m^A$.
(E--F) The distribution of whether yearly citations have fallen
to half of maximum. (G--H) CDF of $\tau$ given $I = 1$ and
$I = 0$.}
\label{fig:ct-para}
\end{figure}

We calculated the introduced parameters for each paper in our cohort.
Fig.~\ref{fig:ct-para} presents the distributions of these parameters across
all papers, allowing us to probe overall patterns of their citation dynamics.
First, Fig.~\ref{fig:ct-para}A, which shows the cumulative distribution
of $t_f^P$ and $t_f^A$, indicates that almost all papers obtained their first
paper citation during $5$ years after publication, while only $30\%$ of papers
got cited by patents in $5$ years. In general, we observe from
Fig.~\ref{fig:ct-para}B that first patent citation occurred after first paper
citation was obtained ($\Delta t_f > 0$) for almost all papers, and the median
lag is $7$ years.

Focusing on $t_m$, Fig.~\ref{fig:ct-para}C indicates that the median number of
years taken to reach maximum yearly citations is $4$ and $8$ years for paper
and patent citations, respectively. Fig.~\ref{fig:ct-para}D shows that the
majority ($78.9\%$) of papers obtained maximum yearly patent citations after
the same event happened to paper citations ($\Delta t_m < 0$), and median lag
is $6$ years for those papers.

Focusing on how citations decrease after the peak, for only $0.6\%$ papers,
their patent citations have not dropped below half of the maximum
($I^P = 0$; Fig.~\ref{fig:ct-para}E); and $9\%$ for the paper citation case
($I^A = 0$; Fig.~\ref{fig:ct-para}F). Fig.~\ref{fig:ct-para}H indicates that
most papers belonging to this category obtain maximum citations very
recently---within $3$ and $7$ years close to the end of the observation period
for the patent and paper citation case, respectively.

For the remaining $99.4\%$ and $91\%$ papers whose yearly patent and paper
citations have decreased below $c_{t_m}/2$, Fig.~\ref{fig:ct-para}G shows that
such decay is very rapid for both $c_t^P$ and $c_t^A$, taking less than $3$ and
$8$ years for patent and paper citations, respectively.

\section{Discussion}

The increasing availability of large-scale datasets that systematically record
how scholarly papers are referenced and mentioned in different channels has
opened new possibilities for searching for the broader impact of research
beyond the traditional scientific community. Previous studies have focused on
news media, clinical guidelines, policy documents, and the social Web. In this
work, we looked at another domain---patented technologies---that has received
little attention so far in the literature, and studied the technological impact
of papers.

Based on a newly-created dataset that links millions of non-patent references
made by U.S. patents to MEDLINE papers, we compared citation statistics derived
from patent references with traditional citations from papers. We found that
only a small fraction---$4\%$---of papers ever got cited by patents. These
papers are mainly from Biochemistry, Molecular Biology, Allergy and Immunology,
Cell Biology, and Chemistry, and are published in leading biomedical journals.
For these papers, there is a positive correlation between the number of patent
and paper citations, although the magnitude is low, leaving much variations to
be explained by other factors. The comparison between the curves of yearly
patent and paper citations reveals that the majority of papers got their first
and maximum paper citation before obtaining patent citations, highlighting
different life-cycles of citation dynamics.

Future work is needed to uncover factors that explain the difference between
patent and paper citations and examine more closely the context of patent
citations. E.g., what are the technological classes of citing patents and what
are the fields of cited papers? To what extent publicly-financed papers are
cited by private-sector patents? Answers to these questions would contribute
further to the understanding of the technological impact of scientific
research.

\section*{Acknowledgments}
I thank the anonymous referees for helpful comments and suggestions and Aditya
Tandon for helping with retrieving Web of Science data, which is provided by
the Indiana University Network Science Institute.

\appendix

\setcounter{figure}{0}
\makeatletter 
\renewcommand{\thefigure}{A.\@arabic\c@figure}
\makeatother

\setcounter{table}{0}
\makeatletter 
\renewcommand{\thetable}{A.\@arabic\c@table}
\makeatother

\section{}

\begin{figure}[t]
\centering
\includegraphics[trim=0mm 5mm 0mm 0mm, width=\columnwidth]{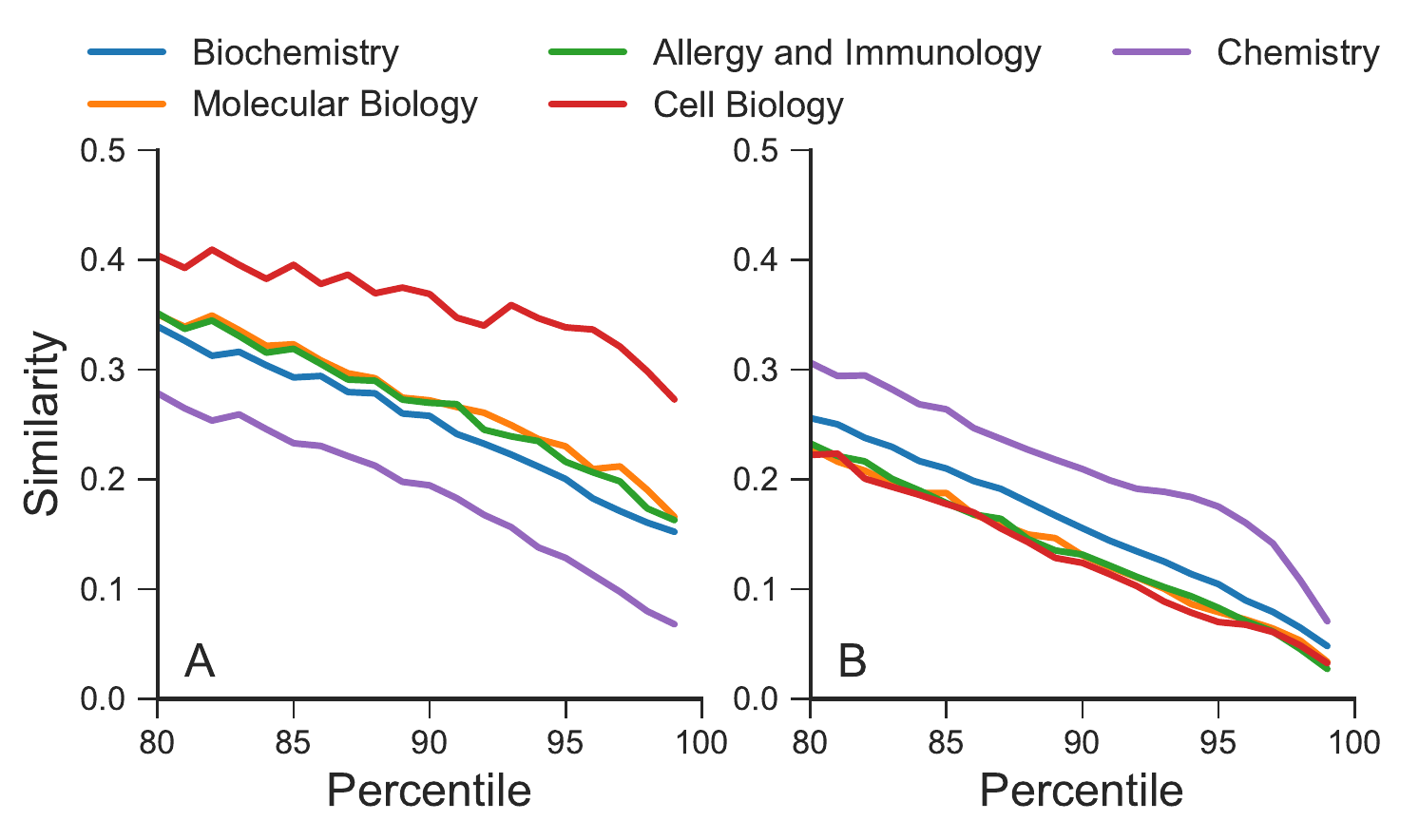}
\caption{Similarity between top papers in different fields. (A) top papers by
$C^P$ and $C^A$. (B) top papers by $B^P$ and $B^A$.}
\label{fig:top-sim}
\end{figure}

\begin{table*}[t]
\tiny
\centering
\caption{Field-specific rankings of top $10$ journals by the total number of
patent citations. We focus on the top $6$ most cited fields listed in
Table~\ref{tab:top-cat}. For each journal, we report the same set of statistics
as in Table~\ref{tab:top-cat}.}
\label{tab:top-jnl}
\begin{tabular}{l r r r r r || l r r r r r}
\hline
\multicolumn{6}{c||}{Biochemistry} & \multicolumn{6}{c}{Science} \\ \hline
Journal & \# C & \% C & \# P & \% P & \% Pub. & Journal & \# C & \% C & \# P & \% P & \% Pub. \\ \hline
\emph{J. Biol. Chem} & $146,369$ & 24.26 & $34,403$ & 23.99 & 20.73 & \emph{PNAS} & $211,529$ & 42.66 & $27,268$ & 47.79 & 23.72 \\
\emph{Nucleic Acids Res.} & $62,952$ & 10.43 & $8,165$ & 5.69 & 22.12 & \emph{Science} & $146,331$ & 29.51 & $12,378$ & 21.70 & 7.60 \\
\emph{Biochemistry} & $46,052$ & 7.63 & $9,795$ & 6.83 & 16.05 & \emph{Nature} & $117,947$ & 23.79 & $11,508$ & 20.17 & 11.53 \\
\emph{Biochem. Biophys. Res. Commun.} & $35,590$ & 5.90 & $9,013$ & 6.29 & 12.00 & \emph{Ann. N. Y. Acad. Sci.} & $10,359$ & 2.09 & $3,372$ & 5.91 & 7.15 \\
\emph{Biochim. Biophys. Acta} & $25,599$ & 4.24 & $7,443$ & 5.19 & 8.08 & \emph{Scientific American} & $2,672$ & 0.54 & 407 & 0.71 & 7.87 \\
\emph{Anal. Biochem.} & $23,093$ & 3.83 & $4,548$ & 3.17 & 18.17 & \emph{Nat. Mater.} & $1,958$ & 0.39 & 318 & 0.56 & 12.55 \\
\emph{FEBS Lett.} & $22,607$ & 3.75 & $5,803$ & 4.05 & 13.42 & \emph{Experientia} & $1,630$ & 0.33 & 589 & 1.03 & 2.88 \\
\emph{Biochem. J.} & $17,935$ & 2.97 & $5,318$ & 3.71 & 9.91 & \emph{Clinical Science} & $1,184$ & 0.24 & 395 & 0.69 & 5.60 \\
\emph{Methods Enzymol.} & $17,172$ & 2.85 & $2,893$ & 2.02 & 15.79 & \emph{Die Naturwissenschaften} & 494 & 0.10 & 105 & 0.18 & 2.06 \\
\emph{FEBS J.} & $15,671$ & 2.60 & $4,305$ & 3.00 & 16.01 & \emph{PloS ONE} & 336 & 0.07 & 236 & 0.41 & 0.29 \\
\hline \hline
\multicolumn{6}{c||}{Molecular Biology} & \multicolumn{6}{c}{Allergy and Immunology} \\ \hline
Journal & \# C & \% C & \# P & \% P & \% Pub. & Journal & \# C & \% C & \# P & \% P & \% Pub. \\ \hline
\emph{EMBO J.} & $30,232$ & 10.72 & $4,990$ & 7.94 & 29.59 & \emph{J. Immunol.} & $54,523$ & 22.43 & $11,903$ & 20.22 & 19.82 \\
\emph{Mol. Cell. Biol.} & $26,616$ & 9.43 & $5,241$ & 8.34 & 24.78 & \emph{J. Exp. Med.} & $29,478$ & 12.13 & $5,002$ & 8.50 & 22.59 \\
\emph{Gene} & $25,893$ & 9.18 & $4,206$ & 6.69 & 24.91 & \emph{Infect. Immun.} & $24,129$ & 9.93 & $6,462$ & 10.98 & 22.07 \\
\emph{J. Mol. Biol.} & $24,384$ & 8.64 & $3,991$ & 6.35 & 13.56 & \emph{J. Immunol. Methods} & $12,066$ & 4.96 & $2,651$ & 4.50 & 25.14 \\
\emph{Nat. Med.} & $12,958$ & 4.59 & $1,795$ & 2.86 & 23.04 & \emph{Eur. J. Immunol} & $10,398$ & 4.28 & $2,407$ & 4.09 & 17.38 \\
\emph{Oncogene} & $11,661$ & 4.13 & $3,187$ & 5.07 & 19.70 & \emph{Vaccine} & $8,782$ & 3.61 & $2,361$ & 4.01 & 14.88 \\
\emph{Genes Dev.} & $9,993$ & 3.54 & $1,892$ & 3.01 & 27.42 & \emph{Mol. Immunol.} & $5,911$ & 2.43 & $1,086$ & 1.84 & 16.72 \\
\emph{Plant Molecular Biology} & $8,593$ & 3.05 & $1,781$ & 2.83 & 28.14 & \emph{Immunol. Today} & $4,871$ & 2.00 & 681 & 1.16 & 23.23 \\
\emph{Mol. Pharmacol.} & $6,549$ & 2.32 & $1,741$ & 2.77 & 16.39 & \emph{Annu. Rev. Immunol.} & $4,827$ & 1.99 & 430 & 0.73 & 53.95 \\
\emph{Mol. Microbiol.} & $6,452$ & 2.29 & $1,846$ & 2.94 & 16.41 & \emph{Immunology} & $4,068$ & 1.67 & $1,267$ & 2.15 & 10.96 \\
\hline \hline
\multicolumn{6}{c||}{Cell Biology} & \multicolumn{6}{c}{Chemistry} \\ \hline
Journal & \# C & \% C & \# P & \% P & \% Pub. & Journal & \# C & \% C & \# P & \% P & \% Pub. \\ \hline
\emph{Cell} & $61,696$ & 31.13 & $6,265$ & 14.92 & 37.77 & \emph{J. Med. Chem.} & $69,438$ & 39.56 & $10,795$ & 24.51 & 42.95 \\
\emph{Mol. Cell. Biol.} & $26,616$ & 13.43 & $5,241$ & 12.48 & 24.78 & \emph{J. Am. Chem. Soc.} & $15,990$ & 9.11 & $4,996$ & 11.34 & 10.96 \\
\emph{J. Cell Biol.} & $16,627$ & 8.39 & $3,744$ & 8.91 & 18.63 & \emph{Bioorganic Med. Chem. Lett.} & $12,341$ & 7.03 & $3,239$ & 7.35 & 17.92 \\
\emph{Plant Cell} & $8,021$ & 4.05 & $1,383$ & 3.29 & 27.59 & \emph{J. Org. Chem.} & $9,716$ & 5.54 & $2,993$ & 6.80 & 13.98 \\
\emph{Exp. Cell Res.} & $5,527$ & 2.79 & $1,735$ & 4.13 & 8.72 & \emph{Chem. Pharm. Bull.} & $7,306$ & 4.16 & $2,009$ & 4.56 & 14.26 \\
\emph{Plant Journal} & $5,185$ & 2.62 & $1,117$ & 2.66 & 19.59 & \emph{J. Neurochem.} & $6,328$ & 3.60 & $2,135$ & 4.85 & 8.61 \\
\emph{J. Cell. Physiol.} & $3,833$ & 1.93 & $1,153$ & 2.75 & 9.88 & \emph{Angewandte Chemie} & $5,168$ & 2.94 & $1,740$ & 3.95 & 8.49 \\
\emph{J. Cell Sci.} & $3,822$ & 1.93 & $1,273$ & 3.03 & 9.78 & \emph{Bioorganic Med. Chem.} & $5,083$ & 2.90 & $1,509$ & 3.43 & 13.74 \\
\emph{Cytometry} & $3,186$ & 1.61 & 737 & 1.75 & 26.26 & \emph{Chem. Rev.} & $4,663$ & 2.66 & 630 & 1.43 & 25.70 \\
\emph{J. Cell. Biochem.} & $2,865$ & 1.45 & 811 & 1.93 & 10.67 & \emph{Chemistry $\&$ Biology} & $3,926$ & 2.24 & 649 & 1.47 & 22.31 \\
\hline
\end{tabular}
\end{table*}

\begin{table*}
\tiny
\centering
\caption{Top $10$ papers by total number of citations from USPTO-issued patents.
For brevity, the ``Author'' column only lists the first author.}
\label{tab:top-ptc}
\begin{tabular}{c r r c l c c}
PMID & $C^P$ & $C^A$ & Author & Title & Year & Journal \\ \hline
2315699 & 2748 & 385 & JU Bowie & Deciphering the message in protein sequences: tolerance to amino & 1990 & \emph{Science} \\
& & & & acid substitutions & & \\
8844166 & 2014 & 125 & Y Eshed & Less-than-additive epistatic interactions of quantitative trait loci & 1996 & \emph{Genetics} \\
& & & & in tomato & & \\
3285178 & 1938 & 44 & E Lazar & Transforming growth factor alpha: mutation of aspartic acid 47 & 1988 & \emph{Mol. Cell. Biol.} \\
& & & & and leucine 48 results in different biological activities & & \\
1699952 & 1770 & 92 & WH Burgess & Possible dissociation of the heparin-binding and mitogenic & 1990 & \emph{J. Cell Biol.} \\
& & & & activities of heparin-binding (acidic fibroblast) growth factor-1 & & \\
& & & & from its receptor-binding activities by site-directed mutagenesis & & \\
& & & & of a single lysine residue & & \\
1172191 & 1749 & 11000 & G K\"ohler & Continuous cultures of fused cells secreting antibody of predefined & 1975 & \emph{Nature} \\
& & & & specificity & & \\
9305836 & 1604 & 1053 & IM Verma & Gene therapy -- promises, problems and prospects & 1997 & \emph{Nature} \\
10631780 & 1499 & 78 & J Skolnick & From genes to protein structure and function: novel applications & 2000 & \emph{Trends Biotechnol.} \\
& & & & of computational approaches in the genomic era & & \\
11325474 & 1473 & 276 & SR Vippagunta & Crystalline solids & 2001 & \emph{Adv. Drug Deliv. Rev.} \\
2271534 & 1266 & 506 & JA Wells & Additivity of mutational effects in proteins & 1990 & \emph{Biochemistry} \\
2231712 & 1235 & 35800 & SF Altschul & Basic local alignment search tool & 1990 & \emph{J. Mol. Biol.} \\
\end{tabular}
\end{table*}

\begin{table*}
\tiny
\centering
\caption{Top $10$ papers by total number of citations from papers.}
\label{tab:top-ppc}
\begin{tabular}{c r r c l c c}
PMID & $C^P$ & $C^A$ & Author & Title & Year & Journal \\ \hline
14907713 & 247 & 275527 & OH Lowry & Protein measurement with the Folin phenol reagent & 1951 & \emph{J. Biol. Chem.} \\
5432063 & 942 & 198137 & UK Laemmli & Cleavage of structural proteins during the assembly of the head of & 1970 & \emph{Nature} \\
& & & & bacteriophage T4 & & \\
942051 & 568 & 145074 & MM Bradford & A rapid and sensitive method for the quantitation of microgram & 1976 & \emph{Anal. Biochem.} \\
& & & & quantities of protein utilizing the principle of protein-dye binding & & \\
271968 & 1164 & 62573 & F Sanger & DNA sequencing with chain-terminating inhibitors & 1977 & \emph{PNAS} \\
2440339 & 569 & 59607 & P Chomczynski & Single-step method of RNA isolation by acid guanidinium & 1987 & \emph{Anal. Biochem.} \\
& & & & thiocyanate-phenol-chloroform extraction & & \\
388439 & 360 & 50513 & H Towbin & Electrophoretic transfer of proteins from polyacrylamide gels to & 1979 & \emph{PNAS} \\
& & & & nitrocellulose sheets: procedure and some applications & & \\
13428781 & 51 & 41232 & J Folch & A simple method for the isolation and purification of total lipides & 1957 & \emph{J. Biol. Chem.} \\
& & & & from animal tissues & & \\
7984417 & 158 & 38418 & JD Thompson & CLUSTAL W: improving the sensitivity of progressive multiple & 1994 & \emph{Nucleic Acids Res.} \\
& & & & sequence alignment through sequence weighting, position-specific & & \\
& & & & gap penalties and weight matrix choice & & \\
2231712 & 1235 & 35800 & SF Altschul & Basic local alignment search tool & 1990 & \emph{J. Mol. Biol.} \\
18156677 & 2 & 34459 & GM Sheldrick & A short history of SHELX & 2008 & \emph{Acta Crystallogr. A} \\
\end{tabular}
\end{table*}

\begin{table*}
\tiny
\centering
\caption{Top $10$ delayed-recognition papers measured based on patent citations.}
\label{tab:top-sb-pt}
\begin{tabular}{c r r r r l l c c}
PMID & $B^P$ & $B^A$ & $C^P$ & $C^A$ & Author & Title & Year & Journal \\ \hline
6804947 & 1244 & 0 & 1180 & 96 & S Rudikoff & Single amino acid substitution altering antigen-binding specificity & 1982 & \emph{PNAS} \\
1353878 & 1084 & 2 & 318 & 10 & PC Huettner & Neu oncogene expression in ovarian tumors: a quantitative study & 1992 & \emph{Mod. Pathol.} \\
8404593 & 1013 & 1 & 325 & 28 & B Freyschuss & Induction of the estrogen receptor by growth hormone and & 1993 & \emph{Endocrinology} \\
& & & & & & glucocorticoid substitution in primary cultures of rat hepatocytes & & \\
1463873 & 993 & 0 & 294 & 6 & E H\"ahnel & Expression of the pS2 gene in breast tissues assessed by & 1992 & \emph{Breast Cancer Res. Treat.} \\
& & & & & & pS2-mRNA analysis and pS2-protein radioimmunoassay & & \\
7507349 & 981 & 4 & 314 & 17 & E Jacquemin & Developmental regulation of acidic fibroblast growth factor & 1993 & \emph{Int. J. Dev. Biol.} \\
& & & & & & (aFGF) expression in bovine retina & & \\
833720 & 939 & 116 & 1012 & 211 & SM Berge & Pharmaceutical salts & 1977 & \emph{J. Pharm. Sci.} \\
8048062 & 935 & 0 & 323 & 82 & ME Hahn & Regulation of cytochrome P4501A1 in teleosts: sustained induction & 1994 & \emph{Toxicol. Appl. Pharmacol.} \\
& & & & & & of CYP1A1 mRNA, protein, and catalytic activity by & & \\
& & & & & & 2,3,7,8-tetrachlorodibenzofuran in the marine fish Stenotomus chrysops & & \\
8275492 & 922 & 0 & 322 & 172 & I Husain & Elevation of topoisomerase I messenger RNA, protein, and catalytic & 1994 & \emph{Cancer Res.} \\
& & & & & & activity in human tumors: demonstration of tumor-type specificity & & \\
& & & & & & and implications for cancer chemotherapy & & \\
7741759 & 856 & 1 & 325 & 64 & J George & Pre-translational regulation of cytochrome P450 genes is responsible & 1995 & \emph{Biochem. Pharmacol.} \\
& & & & & & for disease-specific changes of individual P450 enzymes among & & \\
& & & & & & patients with cirrhosis & & \\
6136691 & 808 & 30 & 131 & 16 & P Svedman & Irrigation treatment of leg ulcers & 1983 & \emph{Lancet} \\
\end{tabular}
\end{table*}

\begin{table*}
\tiny
\centering
\caption{Top $10$ delayed-recognition papers measured based on paper citations.}
\label{tab:top-sb-pp}
\begin{tabular}{c r r r r l l c c}
PMID & $B^P$ & $B^A$ & $C^P$ & $C^A$ & Author & Title & Year & Journal \\ \hline
3899825 & 47 & 553 & 18 & 11308 & DR Matthews & Homeostasis model assessment: insulin resistance and beta-cell & 1985 & \emph{Diabetologia} \\
& & & & & & function from fasting plasma glucose and insulin concentrations & & \\
& & & & & & in man & & \\
728692 & 32 & 464 & 3 & 2683 & RC Young & A rating scale for mania: reliability, validity and sensitivity & 1978 & \emph{Br. J. Psychiatry} \\
477100 & 11 & 448 & 2 & 1553 & TA Gruen & ``Modes of failure'' of cemented stem-type femoral components: & 1979 & \emph{Clin. Orthop. Relat. Res.} \\
& & & & & & a radiographic analysis of loosening & & \\
976387 & 27 & 374 & 12 & 658 & AJ Friedenstein & Fibroblast precursors in normal and irradiated mouse & 1976 & \emph{Exp. Hematol.} \\
& & & & & & hematopoietic organs & & \\
4028566 & 13 & 322 & 2 & 1167 & Y Tegner & Rating systems in the evaluation of knee ligament injuries & 1985 & \emph{Clin. Orthop. Relat. Res.} \\
7018783 & 24 & 313 & 27 & 800 & M Jarcho & Calcium phosphate ceramics as hard tissue prosthetics & 1981 & \emph{Clin. Orthop. Relat. Res.} \\
843571 & 31 & 287 & 4 & 13251 & JR Landis & The measurement of observer agreement for categorical data & 1977 & \emph{Biometrics} \\
70454 & 13 & 267 & 2 & 89 & GW Zack & Automatic measurement of sister chromatid exchange frequency & 1977 & \emph{J. Histochem. Cytochem.} \\
776922 & 22 & 241 & 8 & 264 & ML Landsman & Light-absorbing properties, stability, and spectral stabilization & 1976 & \emph{J. Appl. Physiol.} \\
& & & & & & of indocyanine green & & \\
641088 & 20 & 220 & 4 & 569 & GE Lewinnek & Dislocations after total hip-replacement arthroplasties & 1978 & \emph{J. Bone Jt. Surg} \\
\end{tabular}
\end{table*}

\clearpage


\end{document}